\documentclass[10pt,letterpaper]{article}
\usepackage[top=0.85in,left=2.75in,footskip=0.75in]{geometry}

\usepackage{amsmath,amssymb}

\usepackage{changepage}

\usepackage[utf8x]{inputenc}

\usepackage{textcomp,marvosym}

\usepackage{cite}

\usepackage{nameref,hyperref}

\usepackage{microtype}
\DisableLigatures[f]{encoding = *, family = * }

\usepackage[table]{xcolor}

\usepackage{array}

\newcolumntype{+}{!{\vrule width 2pt}}

\newlength\savedwidth

\raggedright
\usepackage[aboveskip=1pt,labelfont=bf,labelsep=period,justification=raggedright,singlelinecheck=off]{caption}

\makeatletter
\renewcommand{\@biblabel}[1]{\quad#1.}
\makeatother

\usepackage{lastpage,fancyhdr,graphicx}
\usepackage{epstopdf}
\fancyheadoffset[L]{2.25in}
\fancyfootoffset[L]{2.25in}
\begin{document}
\vspace*{0.2in}

\begin{flushleft}
{\Large
\textbf\newline{
Time window to constrain the
corner value
of the global seismic-moment distribution
} 
}
\newline
\\
\'Alvaro Corral\textsuperscript{1,2,3,4*},
Isabel Serra\textsuperscript{1}
\\
\bigskip
\textbf{1} 
Centre de Recerca Matem\`atica, Edifici C, Campus Bellaterra,
E-08193 Barcelona, Spain
\\
\textbf{2} 
Departament de Matem\`atiques,
Facultat de Ci\`encies, Universitat Aut\`onoma de Barcelona,
E-08193 Barcelona, Spain
\\
\textbf{3}
Barcelona Graduate School of Mathematics, Edifici C, Campus Bellaterra,
E-08193 Barcelona, Spain
\\
\textbf{4}
Complexity Science Hub Vienna,
Josefst\"adter Stra$\beta$e 39,
1080 Vienna,
Austria
\\
\bigskip

* alvaro.corral@uab.es
%
%
%
%
%
%


\end{flushleft}
\section*{Abstract}
It is well accepted that, at the global scale, the  Gutenberg-Richter (GR) law describing the distribution of earthquake magnitude or seismic moment has to be modified at the tail to properly account for the most extreme events.  It is debated, though, how much additional time of earthquake recording  will be necessary to properly constrain this tail.  Using the   global CMT catalog, we study how three modifications of the GR law that incorporate a corner-value parameter  are compatible with the size of the largest observed earthquake in a given time window.    Current data lead to  a rather large range of parameter values  (e.g., corner magnitude  from 8.6 to 10.2 for the so-called tapered GR distribution). Updating this estimation in the future will strongly depend on the maximum magnitude observed,  but, under reasonable assumptions, the range will be substantially reduced by the end of this century, contrary to claims in previous literature.




\section*{Introduction}

Statistics 
of earthquake occurrence, {in particular of the most extreme events,} must be
a fundamental source to assess seismic hazard
\cite{Mulargia}.
The cornerstone model for describing the {earthquake-size distribution} 
is  
the Gutenberg-Richter (GR) law 
\cite{Utsu_GR,Kagan_book}.
{The original} version of the GR law states that earthquake magnitudes
follow an exponential distribution, 
{and since this} 
is a perfectly ``well-behaved'' distribution,
with all {statistical} moments  {(such as the mean and the standard deviation) being} finite, 
the problem of earthquake sizes {would seem a rather trivial one.}  

However, a physical interpretation of the meaning of the
GR law needs a proper understanding of magnitude.
In fact, magnitude
presents several difficulties as a measure of earthquake size
\cite{Ben_zion_review},
and a true physical quantity
is given instead by seismic moment 
\cite{Kanamori_77,Kanamori_rpp}.
Due to the logarithmic dependence of magnitude on seismic moment,
the GR law for the latter
transforms into
a power-law distribution,
i.e.,
\begin{equation}
f(x) \propto \frac 1 {x^{1+\beta}}, \mbox{ for } a \le x < \infty,
\label{powerlawuno}
\end{equation}
{where} $x$ is seismic moment, $f(x)$ the seismic-moment probability density,
$a$ a lower cut-off below which the power law does not hold
(presumably because of the incompleteness of the  
{considered catalog for small earthquakes}),  
and $1+\beta$ the power-law exponent,  
{which takes} values close to 1.6 or 1.7
(and with the symbol ``$\propto$'' representing proportionality).
%
%
It turns out  
that the solution to the physical interpretation of the GR law
has a price to be paid: the power-law distribution,
when $1+\beta$ is smaller than 2
(which is indeed the case),
is not ``well behaved'', in the sense that
the mean value {of {the} seismic moment} becomes infinite.

{The reason is} 
that, for power-law distributed seismic moments, 
events in the tail of the distribution,
despite having very small probability, bring an enormous contribution
to seismic-moment release 
\cite{Corral_FontClos}, 
and the seismic-moment sample mean does not converge, no matter how large the number of data is,
due to the inapplicability of the law of large numbers to power-law
distributions 
\cite{Corral_csf} 
such as that in Eq. (\ref{powerlawuno}).
In consequence, 
as 
when extended to the whole range of earthquake sizes
the GR law {is unphysical},
the tail of the distribution of seismic moment
must deviate from the GR power-law shape 
\cite{Kagan_gji02}.

Due to scarcity of data, the problem
has to be approached
at a global scale, or at least for a large subset of the global data
(for instance, for subduction zones as a whole 
\cite{Kagan_gji02}).
This approach has been followed by a number of authors
\cite{Kagan_book,Kagan_gji02,Kagan_pageoph99,Godano_pingue,Main_ng,Kagan_tectono10,Bell_Naylor_Main,Corral_Deluca,Zoller_grl,Geist_Parsons_NH}.
%
%
Essentially, a new parameter $M_c>0$ is introduced, providing a scale
for the seismic moment of the {largest}
{(``non-GR'')} earthquakes, in such a way
that for $x \ll M_c$ the GR law can be considered to hold
but for $x \gg M_c$ the distribution clearly departs from this law,
decaying faster than the GR power law.
The values of ${M_c}$ are more easily read in terms
of the corresponding (moment) magnitude $m_c$ 
\cite{Kanamori_77,Kanamori_rpp},
through the formula
$
m_c= 
2  \left( \log_{10} {M_c} - 9.1\right) /3,
\label{mcMc}
$ 
where the seismic moment
is measured in N$\cdot$m. 
%
%
As $m_c$ is sometimes referred to as ``corner magnitude'',
so $M_c$ would be the ``corner seismic moment'' 
\cite{Kagan_gji02}, 
independently of the specific probabilistic model
(in practice, we will use $M_c$ for formulas and $m_c$ for reporting numeric values, 
and both will be referred to as ``corner parameters'' or ``corner values'').

In this article we aim to further clarify to what extent the available observations can constrain ${M_c}$ or ${m_c}$, and how many more earthquakes (and then, how many more years of recording) would be likely necessary to 
yield reasonably precise values of such estimates.
Before proposing a rigorous statistical way to tackle these issues, we will need first to 
assess a previously proposed approach 
\cite{Zoller_grl}.



\section*{Probabilistic models}


We 
define the probabilistic models in terms of the cumulative distribution function, $F(x)$, which gives the probability that the random variable (seismic moment) is equal or smaller than a value $x$.
This description is totally equivalent to the one in terms of the probability 
density, as both functions are related as $f(x)=dF(x)/dx$ and
$F(x)=\int_a^x f(x') dx'$
(at some point we will use also the complementary
cumulative distribution function, $S(x)=1-F(x)$).

The 
distributions of our interests are:

(i) the truncated power-law (TPL) distribution 
\cite{Zoller_grl},
\begin{equation}
F_{tpl}(x) =
\left[1-\left(\frac a {M_c} \right)^\beta \right]^{-1}
\left[1-\left(\frac a x \right)^\beta \right],
\mbox{ for } a \le x \le {M_c};
\label{Ftpl}
\end{equation}

(ii) the 
tapered (Tap) GR law 
\cite{Bell_Naylor_Main,Zoller_grl,Mulargia_Geller},
also called Kagan distribution 
\cite{Vere_Jones_gji},
\begin{equation}
F_{tap}(x)=
1-\left(\frac a x\right)^\beta e^{-(x-a)/{M_c}},
\mbox{ for } a \le x < \infty;
\label{Ftap}
\end{equation}

(iii) the truncated gamma (TrG) distribution 
\cite{Main_ng,Serra_Corral},
\begin{equation}
F_{trg}(x) = 1-\frac
{\Gamma(-\beta,x/{M_c})}
{\Gamma(-\beta,a/{M_c})},
\mbox{ for } a \le x < \infty;
\label{Ftrg}
\end{equation}
with $\Gamma(\gamma,z)= \int_z^\infty x^{\gamma-1} e^{-x}dx$ 
the upper incomplete gamma function, 
defined when $\gamma <0$ only for $z>0$.
All three $F(x)$ are zero for $x<a$ and
$F_{tpl}(x)=1$ for $x\ge M_c$.
The parameter $\beta$ has to be greater than zero, 
except in the TrG model, where it has no restriction.
Of course, $M_c> 0$ and $a> 0$.

The three distributions are graphically depicted in Figs. 
S1-S3 of the supporting information.
Note that for the TPL distribution
$M_c$ is a truncation parameter,
whereas for the Tap and TrG
it is a 
scale parameter (it sets the scale of $F(x)$ in the $x-$axis) 
\cite{Main_ng,Serra_Corral}. 
{Namely,} 
{$f_{tpl}(x)$} 
goes abruptly (discontinuously) to zero at $x=M_c$,
whereas for the other two distributions this point sets the scale
at which the power law transforms smoothly into an exponential decay.
So, the physical meaning of $M_c$
in the TPL is quite different than in the other two models.  
Note also that the 
{TPL} is truncated both from below
and from above (but the adjective refers to the truncation from above,
$x\le {M_c}$),
whereas the {TrG and Tap are} 
truncated only from below ($x \ge a$).
Summarizing, all the considered distributions have two free parameters,
$\beta$ and ${M_c}$ (or $\beta$ and $m_c$), with the value of $a$
fixed by the completeness {of the earthquake catalog}.
In all cases, the limit ${M_c} \rightarrow \infty$ 
yields the usual power-law (PL) distribution 
\cite{Serra_Corral}, 
$F_{pl}(x)=1-(a/x)^\beta$ for $x\ge a$, 
which is equivalent to
Eq. (\ref{powerlawuno}).
Other works have considered different distributions, 
such as the Gumbel in Ref. 
\cite{Lomnitz}, for which the power-law limit is not so clear.

\section*{State of the art}

Several authors have addressed 
the constraining of the value of $M_c$
and related issues.
In particular, Ref. \cite{Zoller_grl} studied 
the TPL and the Tap distributions
(called there GR and MGR, respectively). 
It was
claimed that,
for global seismicity with magnitude above 5.75
(i.e., seismic-moment lower cut-off $a=5.31 \times 10^{17}$ N$\cdot$m),
an enormous amount of data
{would be} 
necessary in order to obtain reliable estimates
of ${M_c}$ or ${m_c}$
(200,000 years are mentioned for the Tap distribution with $m_c\simeq 8.5$).
Reasonable values
proposed previously by other authors
(for instance $m_c\simeq 9$ 
in Ref. \cite{Bell_Naylor_Main}
for the Tap distribution)
{were} 
discarded. 



\begin{figure}[!htbp] 
\includegraphics[width=.90\columnwidth]{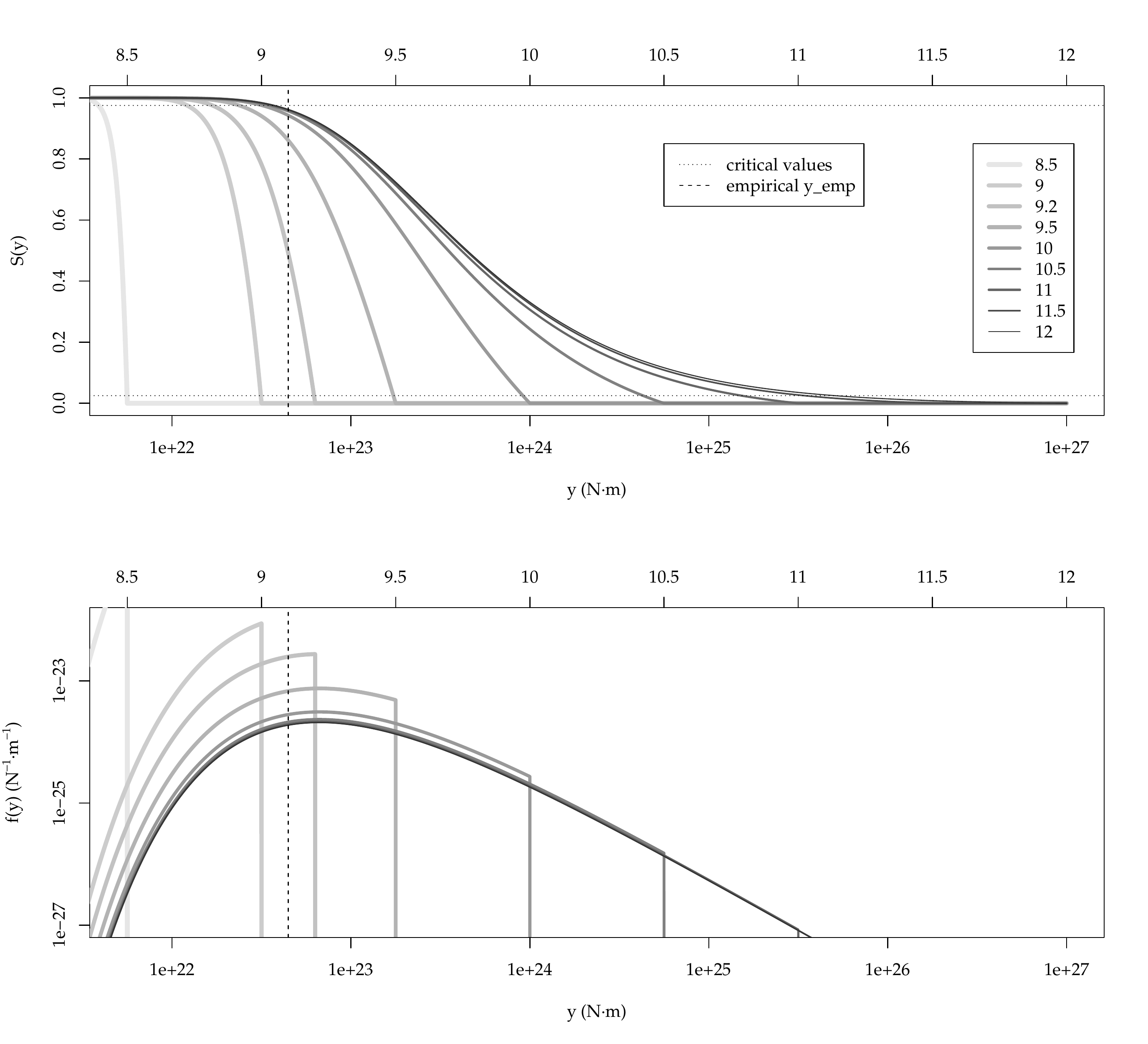}
\caption{
Probability distributions
for the maximum of 
$N=7,585$  {values of seismic moment  (as in the global CMT catalog considered), assuming that these are independent and distributed according to truncated power laws} 
with
lower cut-off $a=5.31 \times 10^{17}$ N$\cdot$m
and diverse values of $m_c$ ranging from 8.5 to 12.
The value of the exponent is fixed to $1+\beta=1.67$,
very close to the maximum-likelihood solution.
The {largest} empirical value {in the catalog}, 
$y_{emp}$, 
is shown as a vertical line.
(a) Complementary cumulative distributions
$S_{max}(y)$
and critical values 0.025 and 0.975 (horizontal lines). 
Note that $0.025 < S_{max}(y_{emp})< 0.975$ at least for $m_c\ge 9.2$,
in contrast with the results of 
Ref. \cite{Zoller_grl},
so these values of $m_c$ cannot be ruled out.
(b) The corresponding probability densities $f_{max}(y)$.
}
\label{figone}
\end{figure}

The analysis 
was based {on a} 
single statistic: 
the maximum {seismic moment} $Y$ of the $N$
{earthquakes with magnitude above 5.75 contained in the catalog}; 
that is,
$$
Y = \max\{X_1,X_2,\dots X_N\}.
$$
Elementary probability theory
allows one {obtaining} 
the {probability} distribution
of the maximum $Y$ when the $N$ observations
are independent 
\cite{Zoller_grl,Ross_firstcourse8}
(independence is the maximum-entropy outcome when there is
no constrain for the dependence between the observations 
\cite{Broderick}).
{Namely,} 
the cumulative distribution function 
of this maximum is given by
\begin{equation}
F_{max}(y) = \mbox{Prob}[ Y \le y ] = [F(y)]^N,
\label{Fmaxx}
\end{equation}
where $F(y)$ 
can be given
by any of the distributions in Eqs. (\ref{Ftpl})-(\ref{Ftrg}),
depending on the underlying statistical model.
{This} 
approach constitutes an ``extreme'' limit
of the classical block-maxima procedure used in extreme-value theory,
considering just one single block 
\cite{Coles}.
Figure \ref{figone} provides
an illustration for $F_{max}(y)$;
Figs. S4-S6 in the supporting information
provide a full picture.

Given a set of $N$ 
observations with empirical maximum
$y_{emp}= \max\{x_1,x_2,\dots x_N\}$
and a modeling probability distribution $F(x)$,
Z\"oller 
\cite{Zoller_grl}
correctly
{argued} 
that, if the data come indeed from $F(x)$,
then, $F_{max}(y_{emp})=\mbox{Prob}[ Y \le y_{emp} ]$
should not be too close to 1{. 
The reason is that}
proximity to 1 would mean that the empirical value $y_{emp}$
is too large in relation to the values of $Y$
that one can expect
from the model distribution $F(x)$ 
{and the number of earthquakes observed}.
Subsequently,
this author
{introduced an \textit{ad-hoc}} 
distinction between
what he called
``not well-sampled'' distributions, characterized by
$F_{max}(y_{emp})=\mbox{Prob}[ Y \le y_{emp} ]$ large (close to 1)
and ``well-sampled'' distributions, characterized by
$F_{max}(y_{emp})$ small{. 
The latter can be} equivalently {characterized} 
by {a large value of the complementary cumulative distribution} 
at $y_{emp}$, that is,
$S_{max}(y_{emp})=1-F_{max}(y_{emp})=
\mbox{Prob}[ Y > y_{emp} ]$ large (close to one).
{In} practice 
{\cite{Zoller_grl}},
\begin{equation}
S_{max}(y_{emp})=
\mbox{Prob}[ Y > y_{emp} ] > 0.99
\label{wellsampledness}
\end{equation}
for ``well-sampled'' distributions 
\cite{Zoller_grl}.
{We} will explain below that this criterion cannot be sustained 
{from a statistical point of view, and will introduce instead a robust criterion}.

Analyzing global data from the centroid moment tensor (CMT) catalog 
\cite{Ekstrom2012,Dziewonski},
from January 1, 1977 to June 30, 2012
(including shallow, intermediate and deep events, $N=7,585$ for $x\ge a$),
Z\"oller 
\cite{Zoller_grl} 
found that the value of the maximum magnitude 
corresponds to {the 2011 Tohoku earthquake, with} magnitude 9.1
(note that the 2004 Sumatra earthquake had a combined multiple-source moment magnitude of 9.3, but only 9.0 with the standard CMT determination 
\cite{Tsai_grl}).
In our work, we will analyze the same dataset, 
for the sake of comparison.
Then, this author \cite{Zoller_grl} 
{evaluated} the performance of the {TPL} and the Tap
distributions 
for different fixed values of the parameter ${M_c}$.
The considered values 
correspond to $m_c=8.5, 9, 9.5, \dots 12$, in addition to $m_c=9.2$.
In contrast, it {was} stated that $\beta$ {was} 
estimated by maximum likelihood for fixed $M_c$.

For the {TPL model}, 
a value of $m_c=9.2$ {resulted} in $\mbox{Prob}[ Y > y_{emp} ]=0.55$
\cite{Zoller_grl},
whereas $m_c=9.5$ and $m_c=10$
led to $\mbox{Prob}[ Y > y_{emp} ]$
very close to one, 
and even closer-to-one values {were} obtained for $m_c \ge 10.5$.
%
%
Following the ``well-sampledness'' criterion,
the value $m_c=9.2$ 
was discarded 
for the TPL model,
despite of having the maximum likelihood
among all the values of the parameters considered,
and values with $m_c \ge 10.5$, with much smaller likelihood, 
{were} preferred. 
However, no preference {was} shown 
between $m_c=10.5$ and any other higher value (for instance $m_c=12$) 
and all the models {were} considered equally likely.
%
%
%
%
For the Tap model, the previous results and the conclusions 
\cite{Zoller_grl}
{were} similar to those for the TPL model, 
and in this way
the value $m_c=9$ (proposed 
in Ref. \cite{Bell_Naylor_Main})
was rejected despite of yielding
maximum likelihood.


The
calculation of the required number of data to perform a
reliable estimation of parameter $M_c$ (or $m_c$)
was obtained by imposing a minimum number of events $N_m$ such that
the distribution becomes ``well-sampled'' \cite{Zoller_grl}, 
in the sense of Eq. (\ref{wellsampledness}).
So, introducing Eq. (\ref{Fmaxx}) into Eq. (\ref{wellsampledness}),
\begin{equation}
S_{max}(y_{emp})=
\mbox{Prob}[ Y > y_{emp} ] = 1 -[F(y_{emp})]^{N_m} > 0.99.
\label{previozoller}
\end{equation}
Note that, no matter the value of $F(y_{emp})$,
if this is 
strictly smaller than 1,
for sufficiently large $N_m$
we will have $[F(y_{emp})]^{N_m} < 0.01$ and the condition will be
fulfilled
by any model, with any parameter value,
if enough data are gathered
(except truncated models with $F(y_{emp})=1$).
%
Imposing that the previous condition becomes an equality one gets
\begin{equation}
N_m=\frac{|\ln 0.01|}{|\ln F(y_{emp})|}=
\frac{7,585 | \ln 0.01|}{|\ln \left(1- \mbox{Prob}[ Y > y_{emp} ]\right)|}.
\label{zollerformula}
\end{equation}
We will argue below that 
this equation (\ref{zollerformula}), used (but not made explicit)
in previous research \cite{Zoller_grl},
does not hold for the problem under consideration.

In this way, for the TPL model with $m_c=9.2$,
accepting the value $\mbox{Prob}[ Y > y_{emp} ]=0.55$,
the
approach just outlined, Ref. \cite{Zoller_grl} [Eq. (\ref{zollerformula}) here], yields that $N_m$ has to be higher than 45,000
(corresponding to 212 years {of earthquake recording}, 
with about 214 earthquakes with $x \ge a$ per year).
For the Tap model with $m_c=8.5$,
for which $\mbox{Prob}[ Y > y_{emp} ]=0.0007$,
one obtains that more than 200,000 years {would be needed}
(from $N_m=50 \times 10^6$, roughly). 
Note the counterintuitive results that 
this
approach leads to:
the larger the corner seismic moment $M_c$,
the less data are required for its estimation,
as contained in Eq. (\ref{zollerformula})
(due to the decrease of $F(y_{emp})$ with $m_c$)
and illustrated for the TPL model in Fig. \ref{numberofearthq}. 

\section*{Proper testing using the maximum seismic moment
}

First, let us show 
that the previously used ``well-sampledness'' criterion 
\cite{Zoller_grl}, 
reproduced here in Eq. (\ref{wellsampledness}), is not appropriate. 
%
If the distribution $F(x)$ is a good model for the empirical data,
what one expects is that both $\mbox{Prob}[ Y \le y_{emp} ]$ and
$\mbox{Prob}[ Y > y_{emp}]$ are not too close to 1,
let us say,
below $1-(1-r) \alpha$ and $1-r \alpha$, respectively,
at significance level $\alpha$
(with $r=1/2$ in the usual symmetric case
and $\alpha=0.05$ or $0.01$).
As both probabilities add to one, the conditions can be written as
\begin{equation}
r\alpha < \mbox{Prob}[ Y \le y_{emp} ] < 1-(1-r)\alpha.
\label{condition_alpha}
\end{equation}
or, equivalently, as
$$
(1-r)\alpha < \mbox{Prob}[ Y > y_{emp} ] < 1-r\alpha,
$$
i.e., the random variable $Y$ takes not too extreme values 
with probability $1-\alpha$ (e.g. 0.95 or 0.99).
Note the profound difference between these conditions
and the ``well-sampledness'' criterion 
\cite{Zoller_grl},
Eq. (\ref{wellsampledness}) here.

Note that, 
following this ``new'' criterion, 
previous numerical results 
for the truncated power-law distribution
\cite{Zoller_grl}
seem to indicate
(in contrast to the conclusions there)
that
all tested values of $m_c$
should be rejected
at the 0.05 significance level
(as Ref. \cite{Zoller_grl} reports
$\mbox{Prob}[ Y > y_{emp} ] > 0.975$),
except $m_c=9.2$ (the value of $\mbox{Prob}[ Y > y_{emp} ]$ for
$m_c=9.5$ displayed in Fig. 3 of 
Ref. \cite{Zoller_grl} 
seems to be
slightly above 0.975 and should be rejected as well,
at least in the symmetric case $r=1/2$).
For the Tap distribution, the only values of $m_c$ that should not be clearly rejected
from 
the numerical results of 
Ref. \cite{Zoller_grl} 
(again in contrast with the conclusions of that reference)
are $m_c=9$ and $m_c=9.2$
(for the rest of $m_c$ values 
Ref. \cite{Zoller_grl} reports
$\mbox{Prob}[ Y > y_{emp} ]$ above 0.975 or below 0.025).
%
%
But 
the numerical results of 
Ref. \cite{Zoller_grl} 
are not in correspondence with ours;
our maximum-likelihood estimations for $\beta$ do not lead to
{$\mbox{Prob}[Y>y_{emp}]\simeq 1$}
when $m_c$ is large ($m_c \ge 10$).
What we find for those values is $\mbox{Prob}[Y>y_{emp}] < 0.975$,
{see} Figs. \ref{figone} and S5
(and Fig. S6 for the TrG),
so all large values of $m_c$ are allowed, in principle.

Regarding the number of earthquakes required to constrain
the corner parameters ($M_c$ or $m_c$),
what is implicit behind Eq. (\ref{zollerformula})
is that a ``not well-sampled'' distribution
(with $\mbox{Prob}[ Y > y_{emp} ]$ close to zero)
is ``not well-sampled'' just because of ``bad luck'',
{that is, the largest earthquake had} $y_{emp}$
much larger than {expected} 
from both the model $F(x)$ and
the actual value of $N$. 
{This} bad luck is what leads to the rejection of the null hypothesis
in usual statistical testing 
(and corresponds to the significance level, see Eq. (\ref{condition_alpha})).
{But}, in
Ref. \cite{Zoller_grl}'s 
argument,
gathering 
more data {would eventually} 
lead to the accommodation
of the theoretical distribution of the maximum
to the empirical value $y_{emp}$, {regardless of the model}.
Thus, in
that
assumption $y_{emp}$ is considered quenched,
i.e., it does not grow despite 
the fact that the number of data increases.
This is hard to justify.

\section*{Proper constraining of the corner seismic-moment: 
TPL case
}

In this section we derive
{a} proper {statistical} way to evaluate the number $N$ of earthquakes necessary to
constrain the estimated value of $M_c$ or $m_c$ for the TPL distribution.
{In this case, our approach uses} 
the distribution 
of the estimator of these quantities ($M_c$ and $m_c$)
to calculate their {statistical} uncertainty as a function of $N$,
and {looks} for the value of $N$ that {reduces} 
{the} uncertainty {down to a desired range}.
{This will necessary} 
depend on the true values of the parameters, which are unknown,
and {is} also {based} on the assumption
that the sample is representative of the whole population
(otherwise, no inference is possible).

For this purpose,
let us {focus} 
in the truncated-power-law model,
which has the peculiar property that the random variable $Y$
(the maximum {seismic moment} of the $N$ {earthquakes}) 
constitutes
the maximum-likelihood
estimator, $\hat M_c$, of the truncation parameter $M_c$, 
{that is} 
$Y=\hat M_c$ for the TPL
(or, equivalently, $\hat m_c$ for the magnitude).
%
%
%
Then, inverting $F_{max}(y_p)=p$,
with $y_p$ defining the $100p-$th percentile of the distribution of the maximum seismic moment
(i.e., the distribution of $\hat M_c$),
one can get the 
probability
of any 
interval for $\hat M_c$.
The limiting points for these intervals are, from 
Eqs. (\ref{Fmaxx}) and (\ref{Ftpl}),
$$
y_{p,tpl}=\frac a{\sqrt[\beta]{1-p^{1/N}[1-(a/M_c)^\beta]}},
$$
and in terms of the magnitude,
\begin{equation}
m_{p,tpl}= \frac 2 3 \left[ \log_{10} \left( \frac a{\sqrt[\beta]{1-p^{1/N}[1-(a/M_c)^\beta]}} \right)- 9.1\right],
\label{mp}
\end{equation}
using 
the relation between 
magnitude and seismic moment,
with $m_{p}$ the $100p-$th percentile of the distribution of the maximum
magnitude.
For the true distribution,
the resulting 95\%-probability
intervals, $(m_{p,tpl},m_{p+0.95, tpl})$,
should contain the empirical value of the maximum with a $0.95$ probability.
These intervals
are shown in Fig. \ref{intervalos}, 
using the empirical value of $N$ {in the global CMT catalog} and different values of $M_c$,
with $\beta$ fixed to $0.67$,
and $p=0.025$ for symmetric intervals
(we have checked that the final results do not depend too much on this choice). 

Figure \ref{intervalos} 
shows that the ideal situation
happens when the distribution of the maximum-likelihood estimator
is very narrow, and then $\hat m_c\simeq m_c$,
leading to the automatic recovering of the true value
(a value very close to it, but below, in fact).
When $N$ is equal to the empirical value 
(considering the case previously studied in the literature \cite{Zoller_grl},
up to mid 2012)
this happens for $m_c<8.5$.
One could refer to this case as {``sampled enough''} 
(in sharp contrast with previous terminology 
\cite{Zoller_grl}).
On the contrary,
when the upper limit of the interval, $m_{p+0.95}$,
departs clearly from the true value of $m_c$,
we may talk of undersampling 
({there is no hint of} the real maximum $m_c$ 
after {the} $N$ observations,
again in contrast with 
previous research
\cite{Zoller_grl}).
This is the case for $m_c > 10.5$ (for $N=7,585$),
for which the intervals do not
include the true value of $m_c$
(for instance, for $m_c=12$ the 
interval
of the maximum
goes from 9 to 11,
roughly, see Fig. \ref{intervalos}). 
But note this kind of undersampling still would allow {ruling}
out the values of the parameters
of the undersampled distributions, if the empirical value of the maximum
were outside the 
resulting
interval
(nevertheless, this is not the case for the actual value, see below).
In the intermediate case
($8.5 < m_c < 10.5$ for the period under consideration),
the intervals are wide but they reach the true value.



\begin{figure}[!htbp] 
\includegraphics[width=.8\columnwidth]{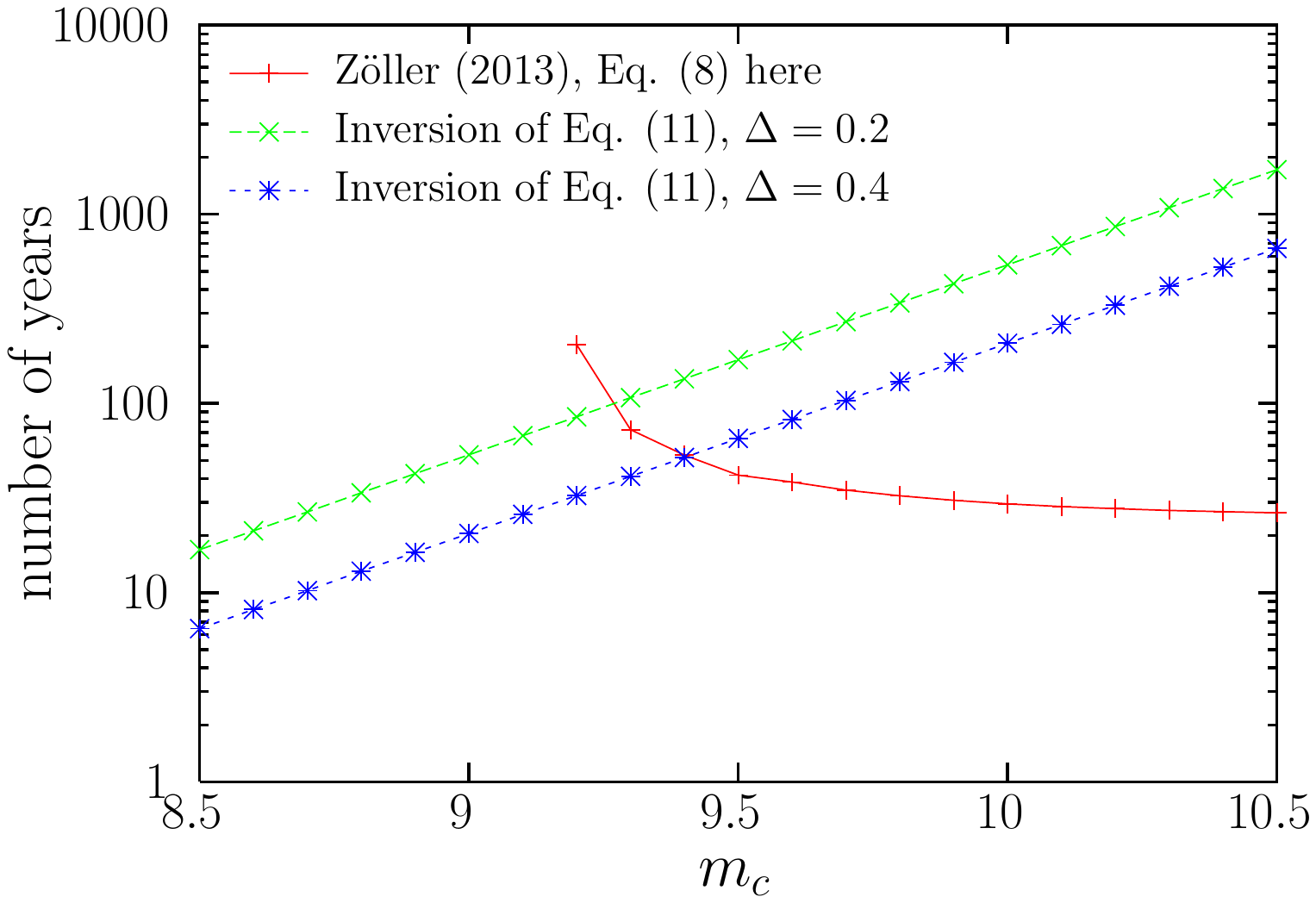}
\centering
\caption{
Number of years necessary to obtain a reliable estimation 
of the truncation parameter $M_c$
for the TPL model with $\beta=0.67$
as a function of the hypothetical true value of $M_c$ (represented by $m_c$),
according to 
Ref. \cite{Zoller_grl} 
(decreasing curve)
and according to our results [inverting Eq. (\ref{ainvertir}), increasing curves],
assuming an average rate of 213.7 events per year.
In the latter case we 
impose that 
95\%-probability
intervals
have magnitude
width $\Delta=0.2$ and $0.4$.
The resulting values of $N$ guarantee
no undersampling
(i.e., $m_{p+0.95}\simeq m_c$, not shown).
Note the totally different outcomes of the two approaches.
}
\label{numberofearthq}
\end{figure}

\begin{figure}[!htbp] 
\includegraphics[width=0.75\columnwidth]{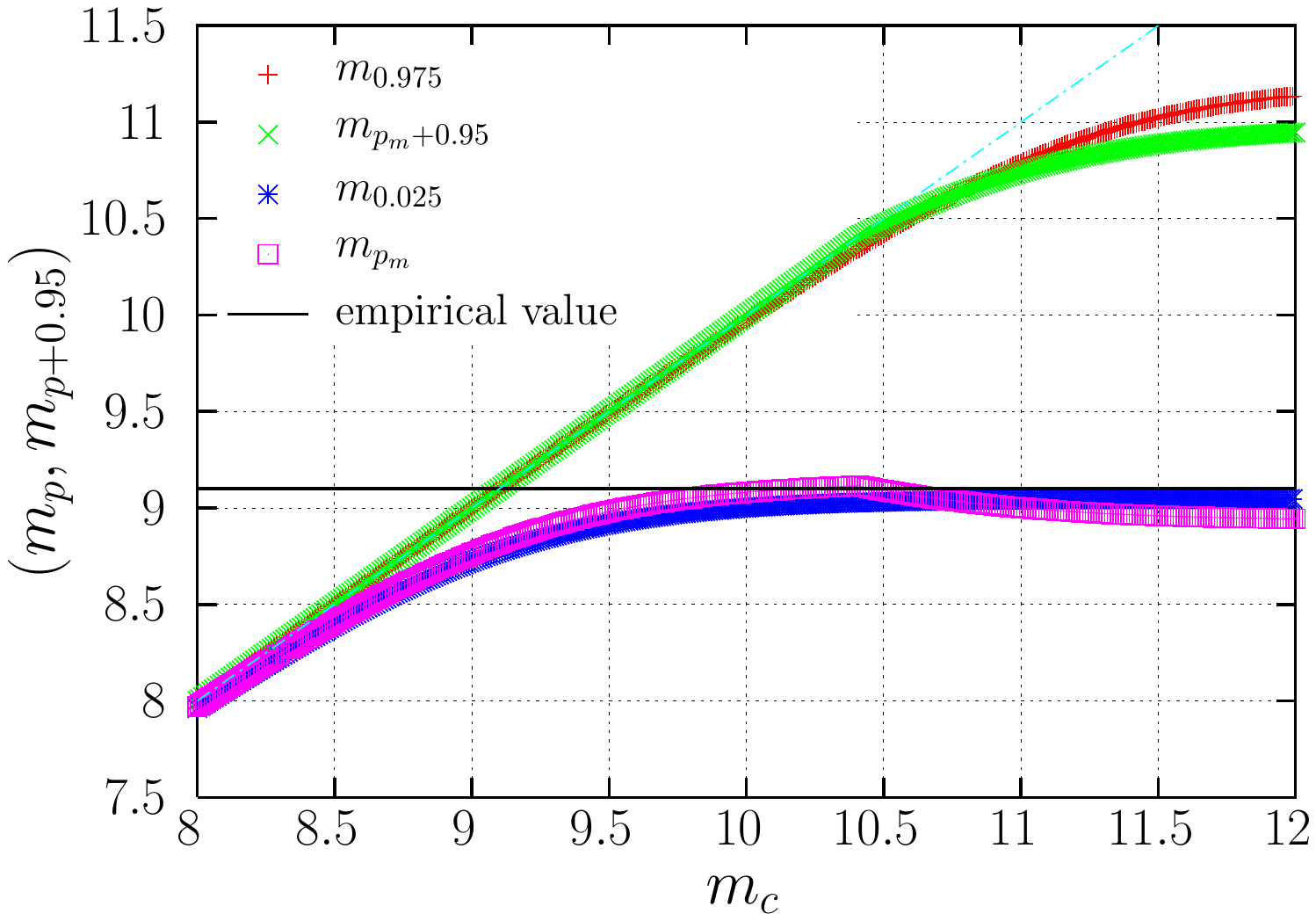}
\caption{
95\%-probability
intervals, 
represented by the starting and ending points
$(m_p,m_{p+0.95})$ 
for 
the truncation parameter $M_c$ of a TPL distribution
with $N=7,585$ {earthquakes} 
(in terms of the corresponding {truncation} magnitude $m_c$),
as a function of
the hypothetical true values of $m_c$.
The value of the exponent is
$1+\beta=1.67$.
Two kinds of 
intervals are shown:
symmetric ($r=1/2$ in Eq. (\ref{condition_alpha}))
and of minimum 
width (the $r$ that gives minimum width is selected),
labeled with $m_{p_m}$.
The empirical value of the maximum observed magnitude 
{in the global CMT catalog}
for the 7,585 considered earthquakes 
is shown as a horizontal line.
When the line is outside the interval, 
the parameter value $m_c$ should be rejected.
}
\label{intervalos}
\end{figure}

%
We can use the previous argument to find the value
of $N$ that leads to narrow 
95\%-probability
intervals for the estimation of $M_c$ or $m_c$
in the TPL model.
Using Eq. (\ref{mp}),
the width of the magnitude intervals,
$\Delta=m_{p+0.95}-m_{p}$,
is obtained as
\begin{equation}
\Delta_{tpl}=
\frac
2 {3\beta}
\log_{10}
{\frac
{1-[1-(a/M_c)^\beta] p^{1/N}}
{1-[1-(a/M_c)^\beta] (p+0.95)^{1/N}}
}.
\label{ainvertir}
\end{equation}
Isolating $N$ as a function of $\Delta_{tpl}$ for given values of $M_c$
and $\beta$ yields the desired result.
Notice that, in contrast to Ref. \cite{Zoller_grl}
[Eq. (\ref{wellsampledness})], our approach 
does not need any empirical information 
(except the value of $\beta$).
Going back to
Fig. \ref{numberofearthq}, this includes
the number of events necessary to obtain
intervals of a fixed width
after numerical inversion of Eq. (\ref{ainvertir}),
as a function of $M_c$.
The results are clearly different to the previous ones 
\cite{Zoller_grl},
as shown in the figure.


Figure \ref{numberofearthq} is {particularly} 
useful {for testing} 
{a specific} value of $m_c$. 
If the real value of $m_c$ were 9.5 (the largest earthquake in the historical record 
\cite{Chile1960}, 
but not contained in the CMT catalog) a 
95\%-probability
interval with width $\Delta=0.4$
(from 9.1 to 9.5, roughly) 
would be obtained after about $N=14,000$
events (corresponding to 65 years, reached in 2042).
If one wants instead a width of $\Delta=0.2$
(yielding an interval from 9.3 to 9.5)
the necessary $N$ is $36,400$, {to be}
reached around the year 2147
({assuming} 
that the TPL were the right model, 
that there is no dependence between the magnitudes,
and that the long-term global earthquake rate and $\beta$ were constant).

It is important to realize
that, in all the cases shown in the figure,
the top value of the interval coincides with the real value.
Although the probability that the estimated value
is between $m_{p+0.95}$ and $m_c$
is $0.05-p$, the two values are very close,
i.e., $m_{p+0.95}\simeq m_c$;
this is due to the extreme sharpness
of the density of the observed maximum close to $m_c$
(for instance, as in Fig. 1, where the vertical axis is logarithmic).
So, the value of $N$ provided in the figure
guarantees no undersampling.
Note also that a 
95\%-probability
interval
is a much more strict requirement
than an interval corresponding to one standard deviation.


We have just calculated the number of earthquakes required
to estimate $M_c$ with a {given} 
uncertainty,
for different hypothetical values of the true $M_c$.
This does not make use the empirical value $y_{emp}$
obtained in $35.5$ years.
A different issue then is how $y_{emp}$ discards or
not the possible values of $M_c$.
Figure \ref{intervalos} shows 
(in addition to the 
intervals of the maximum magnitude obtained from Eq. (\ref{mp}))
the empirical value obtained for the period $1977$-$2012.5$.
If the observed maximum magnitude (9.1 {in the global CMT catalog})
is inside the 
interval, 
there is no reason to reject the parameters
of the model (with a 95\% confidence);
on the contrary, if the empirical value is outside,
we should reject the parameters.



The figure shows how, for the TPL model,
no value of $m_c\ge 9.1$ can be rejected,
i.e., any value of $m_c$ between $9.1$ and $\infty$
is compatible with the empirical result,
and therefore the data do not
allow to determine an upper bound for $m_c$,
although values of $m_c$ above 10 are close to rejection
(with a 95\% confidence; if
we decreased the confidence
or increased the number of data
an upper bound would appear).
%
Indeed, considering the most recent data
at the time of writing, up to the end of 2017
(where no other earthquake of magnitude larger than 9.1 has taken place)
the range of compatible values of $m_c$ turns out to be $9.1$--$10.8$,
as reported in Table \ref{table_data}.

As an illustration, we also analyze what an hypothetical
$y_{emp}$ corresponding to a 9.1 magnitude in a 71-year period
(from 1977 to 2047, let us say) would imply.
Table \ref{table_data} shows
that that would constrain $m_c$ to be between 9.1 and 9.5,
for 
95\%-probability
intervals, but
if the maximum in the same period were 9.3,
the allowed range would be between 9.3 and 10.3.
In contrast, a maximum empirical value of 9.5 (or higher) in that period would yield
$m_c$ unbounded from above again.
Needless to say, we need to wait about 30 years to chose
between these three answers.

%

\section*{Proper constraining of the corner seismic-moment: 
Tap and TrG cases
}

Note that, although the maximum empirical value of the
seismic moment is the maximum-likelihood estimator
of $M_c$
only for the TPL distribution
(out of the three considered models),
we can still use the previous procedure to constrain the value of $M_c$
for any distribution,
but with the resulting values of $M_c$
not related to maximum likelihood estimation, in general. 
Thus, for the Tap distribution,
the percentiles of the maximum seismic moment turn out to be,
using Eq. (\ref{Ftap}) and (\ref{Fmaxx}),
$$
y_{p, tap}=\beta M_c W\left(\frac {a e^{a/(\beta M_c)}}
{\beta M_c (1-p^{1/N})^{1/\beta}}\right),
$$
with $W$ the Lambert W function 
\cite{Corless1996}, 
fulfilling $z=W(z e^z)$.
And for the truncated gamma we get, using Eq. (\ref{Ftrg}),
$$
y_{p, trg}=M_c \Gamma_2^{-1} (-\beta, (1-p^{1/N})\Gamma(-\beta, a/M_c) ),
$$
with $\Gamma_2^{-1}$ the inverse, respect to its second argument,
of the incomplete gamma function.
%
In the same way as 
for the TPL, the empirical value $y_{emp}$
leads to an unbounded range of the values of $m_c$ compatible
with $y_{emp}$ for the original value of $N$ (7,585). These ranges go from $8.65$ to $\infty$ for the Tap distribution and from $8.8$ to $\infty$ for the TrG, with $\beta=0.67$.
However, when one extends the analysis up to 2017 the ranges become bounded,
although large, see Table \ref{table_data}.

This table also explores the values of these ranges in the future,
depending on the hypothetical value of the maximum magnitude observed.
We see that,
in general, the ranges provided by the Tap distribution are
somewhat {wider} 
than those provided by the TPL,
whereas the TrG yields rather larger ranges.
This means that the number of data necessary to constrain
the value of $m_c$ is larger in the TrG than in the other two distributions.
The table also allows us to rule out the scenario that
there will be no earthquakes larger than magnitude 9.1 before 2097
for a TPL distribution, as this scenario leads to the implausibility of having events larger than 9.3,
contrary to what was observed in the 9.5 1960 event in Chile
(although the CMT catalog would probably underestimate the seismic moment of such an event 
\cite{Tsai_grl}).

\section*{Discussion}

Before concluding, we briefly explore the implications
of our results for the assesment of seismic hazard.
Considering as an illustration the case of the tapered model, 
we have seen (Table 1) how the CMT data, up to 2017,
is compatible with a range of values of the corner magnitude, 
from $m_{c\,min}=8.6$ to $m_{c\,max}=10.2$
(with a 95 \% confidence).
Therefore, the resulting seismic-moment distribution
(or, in the same way, the resulting magnitude distribution)
will be a mixture (or combination) of the different $S_{tap}(x|M_c)$
(now we use the complementary cumulative distribution function and
make explicit in the notation the dependence on the corner seismic moment $M_c$), 
with $M_c$ ranging from $M_{c\,min}$ to $M_{c\,max}$.
Thus,
\begin{equation}
S_{mix}(x) =\int_{M_{c\,min}}^{M_{c\,max}}
S_{tpl}(x|M_c) \rho(M_c) d M_c,
\label{laintegral}
\end{equation}
where the resulting distribution $S_{mix}(x)$ is no longer a Tap distribution
but a mixture of Tap's with different $M_c$.
The term $\rho(M_c)$ gives weight to the different values of $M_c$.
The same equation holds
for any other probabilistic model (such as TPL and TrG).

One could assume a uniform distribution of corner magnitudes
(all its values would be equally likelly from $m_{c\,min}$ to $m_{c\,max}$).
Interestingly, 
for the corner seismic-moment distribution, 
this leads to the Jeffreys prior of a scale parameter,
$\rho(M_c) \propto 1/M_c$.
Under this choice, the integral in Eq. (\ref{laintegral})
can be easily evaluated by the Monte-Carlo method.
For the Tap model, the probability of an 
earthquake of magnitude 9.1 or larger  
(among all earthquakes with magnitude larger than 5.75)
turns out to be 
$S_{mix}(x)=2.6 \times 10^{-4}$, corresponding to about
1 in 20 years.
In comparison with the CMT catalog itself 
(1 of such events in 35.5 years)
this probability seems somewhat large.
Even higher values of the magnitude or other models
(TPL or TrG)
also seem to lead to an overestimation of
these probabilities.
Naturally, this is the core problem in the statistics of extreme events,
one has very few extreme events to contrast estimations.
As the result is highly sensitive to the choice of the distribution $\rho(M_c)$, this is a topic that deserves further study.

Our results can also have applications for time-dependent hazard
\cite{Mulargia_Geller}.
If we know when the last earthquake of a given seismic moment $x$ or higher
happened (a time $t$ ago), we can obtain the probability of recurrence in a given time period $\Delta$ from the present as
$$
\mathcal{P}_{x,t,\Delta}=
\mbox{Prob}[t < \mbox{ waiting time } \le t+\Delta \,|
\, \mbox{ waiting time } > t] = 1 - \frac{S_w(t+\Delta \, | \, x)}{S_w(t \, | \,x)},
$$
where the subindex $w$ denotes that the distribution refers to the waiting time (not to the seismic moment).
For a Poisson process $S_w$ is exponential with rate $\lambda_x$ and then we recover
$$\mathcal{P}_{x,t,\Delta}=
1-e^{-\lambda_x \Delta} \simeq \lambda_x \Delta =R_a S(x) \Delta,
$$
which turns out to be independent on $t$ and becomes essentially the same formula used above for time independent hazard, with $R_a=213.7$ year$^{-1}$
(we have assumed $\Delta \ll \lambda_x^{-1}$).

In order to obtain time-dependent hazard one needs to go beyond Poisson occurrence.
At a global scale it has been pointed out that the gamma distribution
can describe well earthquake waiting times \cite{Corral_prl.2004,Corral_calcutta};
nevertheless, for the sake of simplicity, we are going to illustrate
the calculation with the Weibull distribution, 
which can give similar fits \cite{Morina_storms}.
In this way, from the equation ago we can write
\begin{equation}
\mathcal{P}_{x,t,\Delta}=1-\exp\left[\left(\frac{t}{c_x}\right)^\gamma -
\left(\frac{t+\Delta}{c_x}\right)^\gamma \right],
\label{mathcalP}
\end{equation}
with $\gamma$ and $c_x$ the shape and scale parameters 
of the Weibull distribution, respectively
(the latter depending on $x$). 
The Poisson case is included in the particular limit $\gamma=1$.

The scale parameter of the waiting-time distribution
can be directly related to the seismic-moment distribution:
On the one hand, the number of events per unit time (with seismic moment above $x$) is $R_a S(x)$. 
On the other hand, this number is also given by $1/\langle t(x) \rangle$, 
where $\langle t (x)\rangle$ is the mean waiting time for events above $x$.
In the particular case of the Weibull distribution, this is given by
$\langle t(x) \rangle = c_x g(\gamma)$ with $g(\gamma)=\Gamma(1+\gamma^{-1})$.
Thus, 
$$
c_x=\frac 1 {g(\gamma) R_a S(x)},
$$
which substituting into Eq. (\ref{mathcalP})
allows the calculation of the probability $\mathcal{P}_{x,t,\Delta}$.
In the case $\Delta \ll t$ this can be simplified to
$$
\mathcal{P}_{x,t,\Delta} \simeq 1-\exp\left[-\gamma \Delta
\left(g(\gamma) R_a S(x)\right)^\gamma t^{\gamma-1}  \right].
$$
In the context of this article, the seismic-moment distribution $S(x)$
could be substituted by the mixture for different values of $M_c$
given by Eq. (\ref{laintegral}).
Nevertheless,
the calculation of these probabilities needs the accurate fitting of the 
waiting time distributions $S_w(t \, | \, x)$
(i.e., the fitting of $\gamma$ and $c_x$ in the case of the Weibull distribution).
This is left to future works.

\section*{Conclusions}

{Summarizing the main results of the article,
we have reconsidered to what extent the available earthquake record can constrain the parameter that characterizes the tail of the global seismic-moment distribution: 
a corner seismic moment ($M_c$, or its corresponding moment magnitude $m_c$), for three different distributions (truncated power law, tapered GR, and truncated gamma).}
We have corrected some of the drawbacks of 
previous literature,
regarding the number of events necessary {for such a purpose}. 

The key point in our approach is to obtain the percentiles of the distribution of the maximum {seismic moment} of $N$ earthquakes,
and {to derive} from {there} 
probability
intervals 
that can be compared with the {maximum seismic moment} observed, $y_{emp}$. 
If $y_{emp}$ is inside the 
interval 
there is no reason to reject the considered value of the corner parameter.
Although currently (up to the end of 2017), 
the range of values of $m_c$ is rather {wide}, 
in 80 years from now these ranges {are expected to} 
decrease substantially,
but
{depending crucially} 
on the maximum value {to be} observed.
For instance, if this were 9.3, the {tapered} 
model would lead to $m_c \simeq 9.1 \pm 0.3$ (95 \% confidence),
and the {truncated gamma} 
model to $9.35 \pm 0.45$ 
(see Table \ref{table_data} for more hypothetical examples).
From here we conclude that the {much larger} 
periods of time {estimated} 
earlier
are not justified.
In addition, for the same reasons elaborated in this article, 
the standard errors of corner parameters 
that we 
\cite{Serra_Corral} 
calculated previously
for almost 37 years of shallow global seismicity
using asymptotic likelihood theory 
do not provide a convenient description of the range of uncertainty
in those parameters.





\section*{Acknowledgements}

We appreciate the critical reading of \'Alvaro Gonz\'alez, as well as  suggestions from the reviewers.
The data used in this research can be downloaded from 
{\tt http://www.globalcmt.org}
\cite{Ekstrom2012}.


%
%
%



\section{Supporting information}

{\bf S1 Fig.} {ccdf $S(x)$ and pdf $f(x)$ of 
TPL distribution with $\beta=0.67$, $a$ corresponding to
moment magnitude $5.75$, and
$M_c$ corresponding to the values of $m_c$ shown in the legend.
}

\noindent
{\bf S2 Fig.} {ccdf $S(x)$ and pdf $f(x)$ of 
Tap distribution with $\beta=0.67$, $a$ corresponding to
moment magnitude $5.75$, and
$M_c$ corresponding to the values of $m_c$ shown in the legend.
}

\noindent
{\bf S3 Fig.} {ccdf $S(x)$ and pdf $f(x)$ of 
TrG distribution with $\beta=0.67$, $a$ corresponding to
moment magnitude $5.75$, and
$M_c$ corresponding to the values of $m_c$ shown in the legend.
}

\noindent
{\bf S4 Fig.} {ccdf $S_{max}(y)$ and pdf $f_{max}(y)$ of the maximum of 
7,585 TPL observations with
$\beta=0.67$, $a$ corresponing to moment magnitude $5.75$, and 
$M_c$ corresponding to the values of $m_c$ shown in the legend.
Critical values at the 95\% confidence level are shown as horizontal lines.
Empirical value of maximum seismic moment observed is shown as 
a vertical line.
Note that this is exactly Fig. 1 of the main text, 
repeated here for completeness.
}

\noindent
{\bf S5 Fig.} {ccdf $S_{max}(y)$ and pdf $f_{max}(y)$ of the maximum of 
7,585 Tap observations with
$\beta=0.67$, $a$ corresponing to moment magnitude $5.75$, and $M_c$ corresponding to the values of $m_c$ shown in the legend.
Critical values at the 95\% confidence level are shown as horizontal lines.
Empirical value of maximum seismic moment observed is shown as 
a vertical line.
}

\noindent
{\bf S6 Fig.} {ccdf $S_{max}(y)$ and pdf $f_{max}(y)$ of the maximum of 
7,585 TrG observations with
$\beta=0.67$, $a$ corresponing to moment magnitude $5.75$, and $M_c$ corresponding to the values of $m_c$ shown in the legend.
Critical values at the 95\% confidence level are shown as horizontal lines.
Empirical value of maximum seismic moment observed is shown as 
a vertical line.
}

%

\newpage

\begin{table}[!ht]
\caption
{Values of the corner parameter $m_c$ compatible 
(for 
95\%-probability intervals)
with a maximum observed magnitude $m(y_{emp})$
in a time period starting in 1977 and ending in the indicated final year, for the 
{truncated power law (TPL), tapered (Tap) and truncated gamma (TrG)} 
distributions. 
The values of $m(y_{emp})$ marked with an {asterisk} (*) 
indicate hypothetical values (the rest corresponds to the real observed value, 9.1).
The value of $\beta$ is 0.67.
The final year is estimated assuming a {global rate of} 213.7 earthquakes {with moment magnitude $\ge 5.75$} per year. 
}
\label{table_data}

\begin{tabular}{ c c lll }
\hline
 &  &  $m_c$ &  $m_c$ &  $m_c$\\
final year & $m(y_{emp})$ &  TPL &  Tap &  TrG\\
\hline
2012.5                & 9.1\phantom{$^*$} & $9.1$--$\infty$  &$8.6$--$\infty$&  $8.8$--$\infty$ \\
2017\phantom{.0}& 9.1\phantom{$^*$} & $9.1$--$10.8$    &$8.6$--$10.2$&   $8.8$--$11.2$ \\
2047\phantom{.0}& 9.1$^*$                 & $9.1$--$ ~9.5$  &$8.6$--$ ~9.3$&  $8.7$--$~9.7$ \\
2047\phantom{.0} & 9.3$^*$                & $9.3$--$10.3$    &$8.8$--$~9.95$& $9.0$--$10.6$ \\
2047\phantom{.0} & 9.5$^*$                & $9.5$--$\infty$  &$9.1$--$\infty$&  $9.2$--$\infty$ \\
2097\phantom{.0} & 9.1$^*$                & $9.1$--$ ~9.3$  &$8.6$--$~9.1$&   $8.7$--$~9.4$ \\
2097\phantom{.0} & 9.3$^*$                & $9.3$--$ ~9.6$  &$8.8$--$~9.4$&   $8.9$--$~9.8$ \\
2097\phantom{.0} & 9.5$^*$                & $9.5$--$10.3$    &$9.0$--$10.0$&   $9.2$--$10.6$ \\
\hline
\end{tabular}
\end{table}

\newpage

\noindent
\Large{\bf Supporting Information for ``Time window to constrain the
corner value of the global seismic-moment distribution''}

%
%
%
%
%
\noindent\textbf{Contents of this file}

\begin{enumerate}
\item Text S1
\item Figures S1 to S6
\end{enumerate}

\noindent\textbf{Introduction.} 
We provide 
graphical description of the three models for seismic-moment distributions in the
paper as well as for the maximum of 7,585 of such events.
We provide also the mathematical expressions for the probability density functions (pdf) of
the three considered models for seismic moment, $x$. 
Figures
$S1$, $S2$, and $S3$ display the probability density functions and complementary
cumulative distribution functions (ccdf) for each model when lower
cut-off and shape parameter are fixed, and parameter $M_c$ is varying
(as indicated in the legend). 
Figures $S4$, $S5$,
and $S6$ correspond to the pdf and ccdf of the maximum of 7,585
observations for each of the previous models.
\\

\newpage
\noindent\textbf{Text S1.}

The mathematical expressions of the probability density functions of the three models 
follow:

Truncated power-law (TPL) distribution,
\begin{equation}
f_{tpl}(x) =\left[\frac {1} {1-(a/M_c)^\beta}\right] \frac \beta
a\left(\frac a x\right)^{1+\beta},
\end{equation}
for $a \le x \le M_c$, and zero otherwise, with $\beta > 0$. 

Tapered (Tap) GR distribution, also called Kagan distribution,
\begin{equation}
f_{tap}(x)=\left[\frac \beta a \left(\frac a x\right)^{1+\beta} +
\frac 1 {M_c} \left(\frac a x\right)^{\beta}\right]
e^{-(x-a)/{M_c}},
\end{equation}
for $a \le x < \infty$, and zero otherwise, with $\beta >0$. 

Truncated gamma (TrG) distribution,
\begin{equation}
f_{trg}(x) = \frac 1 {{M_c} \Gamma(-\beta,a/{M_c})} \left(\frac
{M_c} x \right)^{1+\beta} e^{-x/{M_c}},
\end{equation}
for $a \le x < \infty$, and zero otherwise, with $-\infty <\beta <
\infty$ and $\Gamma(\gamma,z)= \int_z^\infty x^{\gamma-1} e^{-x}dx$
{being} the upper incomplete gamma function, defined when $\gamma
<0$ only for $z>0$.

\begin{figure}
\includegraphics[width=\textwidth]{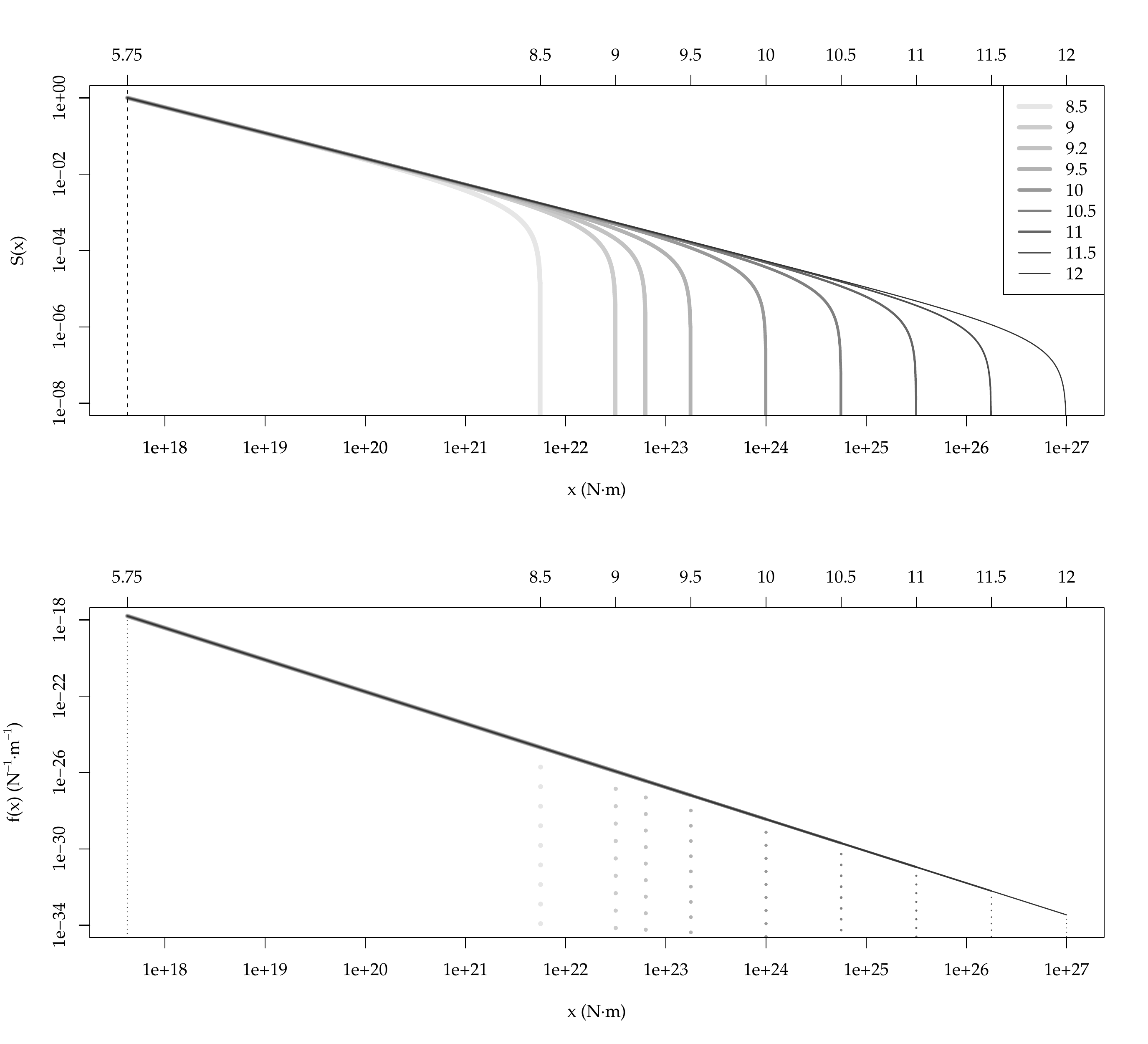}%

\caption{{\bf (Fig. S1)} 
ccdf $S(x)$ and pdf $f(x)$ of 
TPL distribution with $\beta=0.67$, $a$ corresponding to
moment magnitude $5.75$, and
$M_c$ corresponding to the values of $m_c$ shown in the legend.
}
\end{figure}

\begin{figure}
\includegraphics[width=\textwidth]{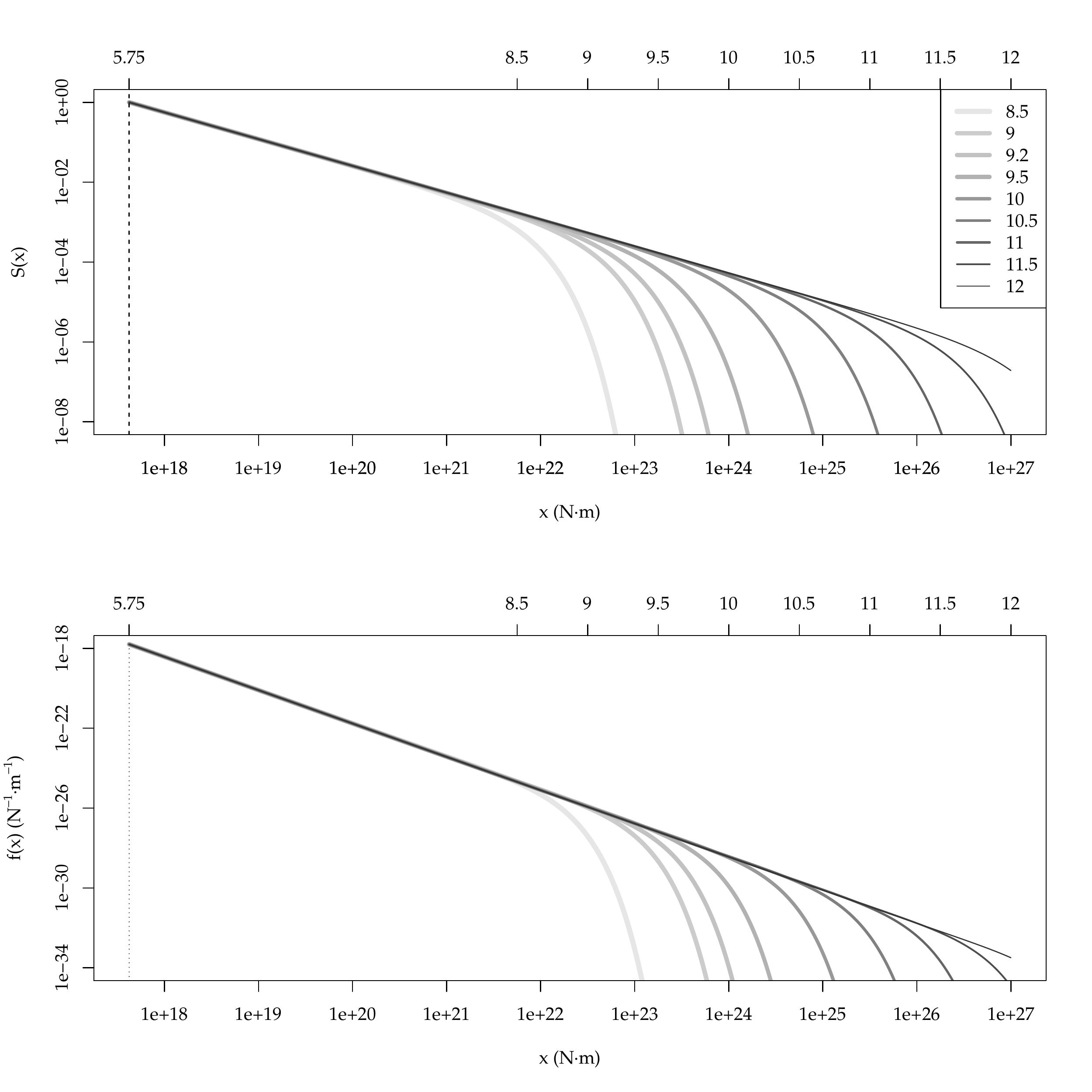}%

\caption{{\bf (Fig. S2)} ccdf $S(x)$ and pdf $f(x)$ of 
Tap distribution with $\beta=0.67$, $a$ corresponding to
moment magnitude $5.75$, and
$M_c$ corresponding to the values of $m_c$ shown in the legend.
}
\end{figure}

\begin{figure}
\includegraphics[width=\textwidth]{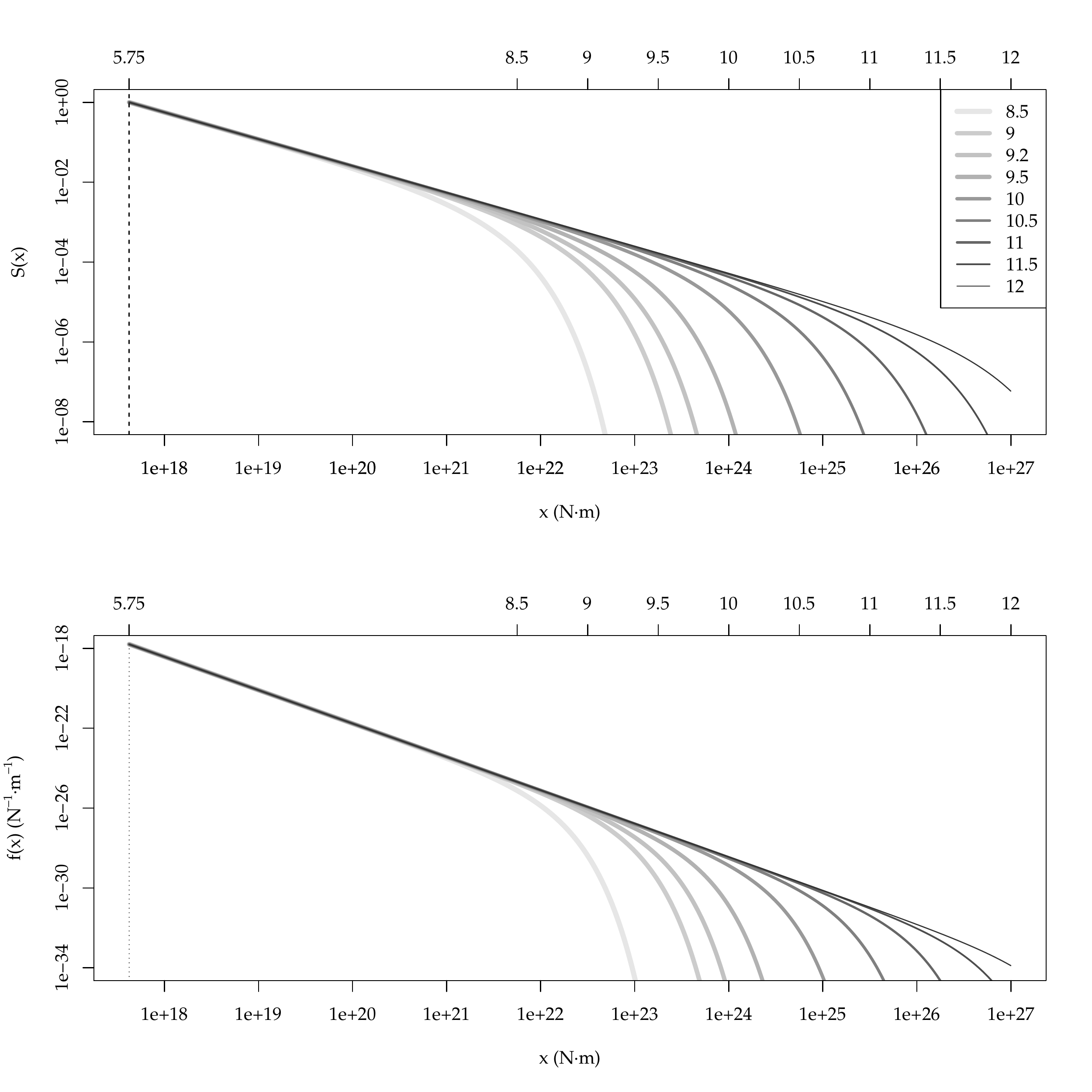}%

\caption{{\bf (Fig. S3)} ccdf $S(x)$ and pdf $f(x)$ of 
TrG distribution with $\beta=0.67$, $a$ corresponding to
moment magnitude $5.75$, and
$M_c$ corresponding to the values of $m_c$ shown in the legend.
}
\end{figure}

\begin{figure}
\includegraphics[width=\textwidth]{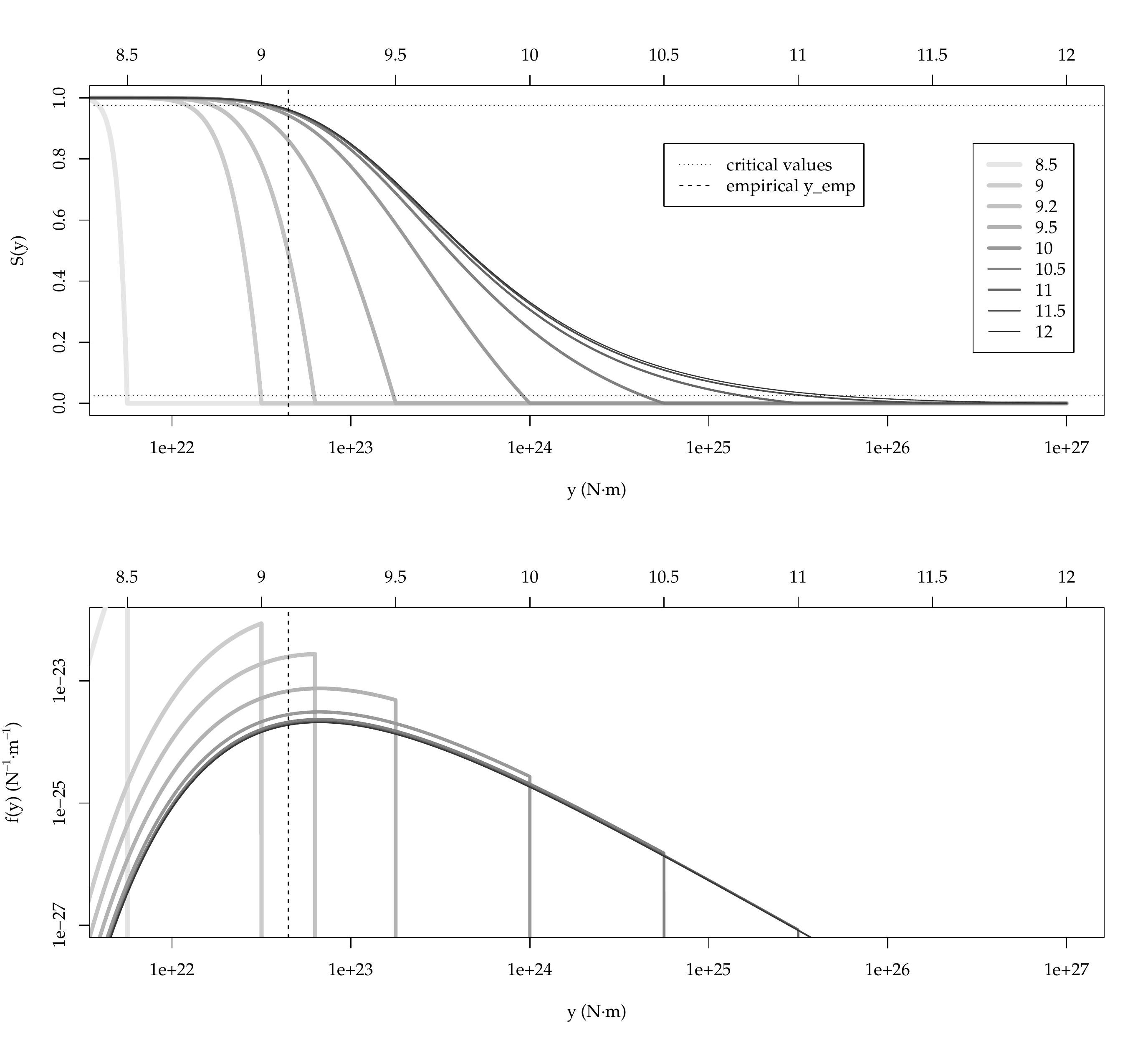}%

\caption{{\bf (Fig. S4)} ccdf $S_{max}(y)$ and pdf $f_{max}(y)$ of the maximum of 
7,585 TPL observations with
$\beta=0.67$, $a$ corresponing to moment magnitude $5.75$, and 
$M_c$ corresponding to the values of $m_c$ shown in the legend.
Critical values at the 95\% confidence level are shown as horizontal lines.
Empirical value of maximum seismic moment observed is shown as 
a vertical line.
Note that this is exactly Fig. 1 of the main text, 
repeated here for completeness.
}
\end{figure}

\begin{figure}
\includegraphics[width=\textwidth]{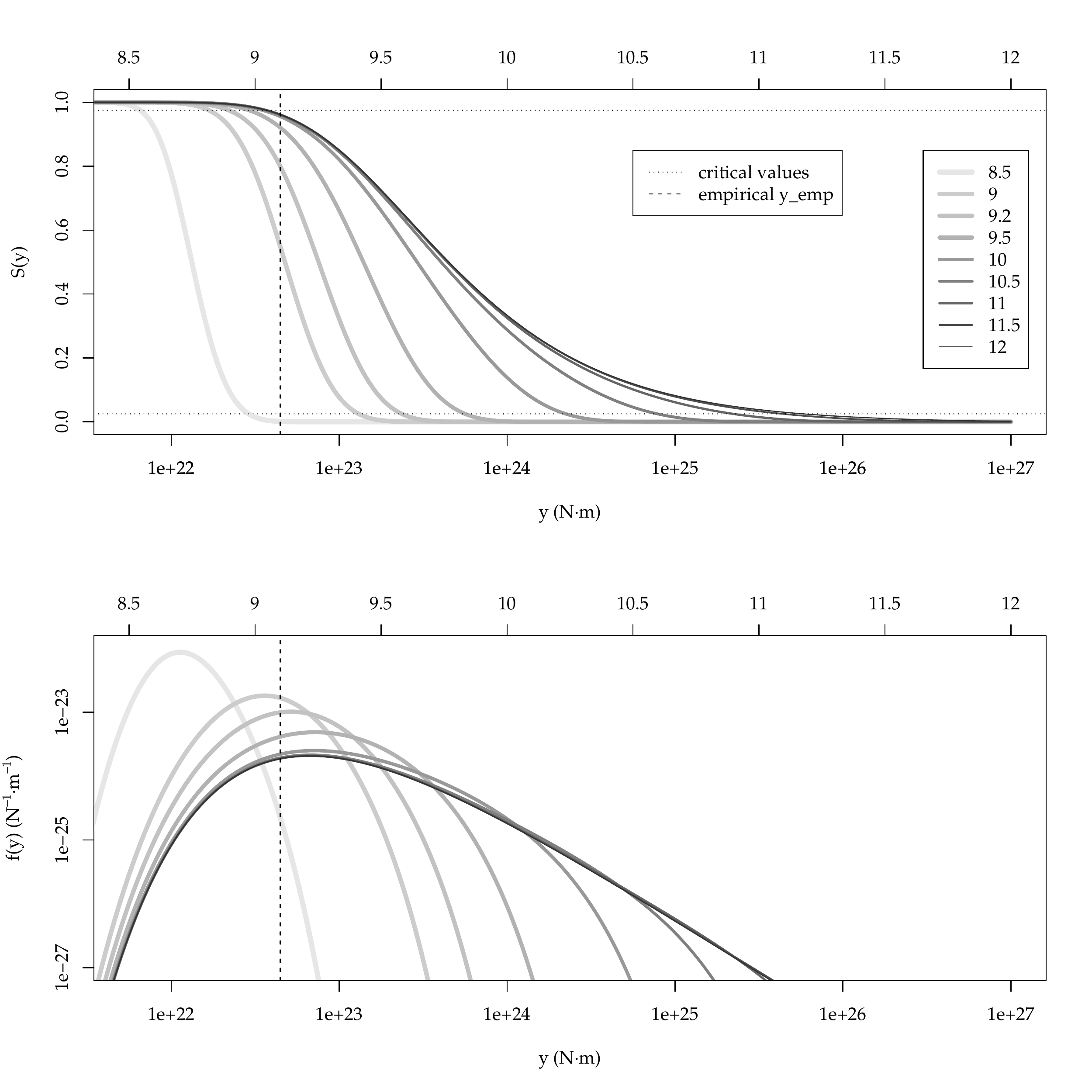}%

\caption{{\bf (Fig. S5)} ccdf $S_{max}(y)$ and pdf $f_{max}(y)$ of the maximum of 
7,585 Tap observations with
$\beta=0.67$, $a$ corresponing to moment magnitude $5.75$, and $M_c$ corresponding to the values of $m_c$ shown in the legend.
Critical values at the 95\% confidence level are shown as horizontal lines.
Empirical value of maximum seismic moment observed is shown as 
a vertical line.
}
\end{figure}

\begin{figure}
\includegraphics[width=\textwidth]{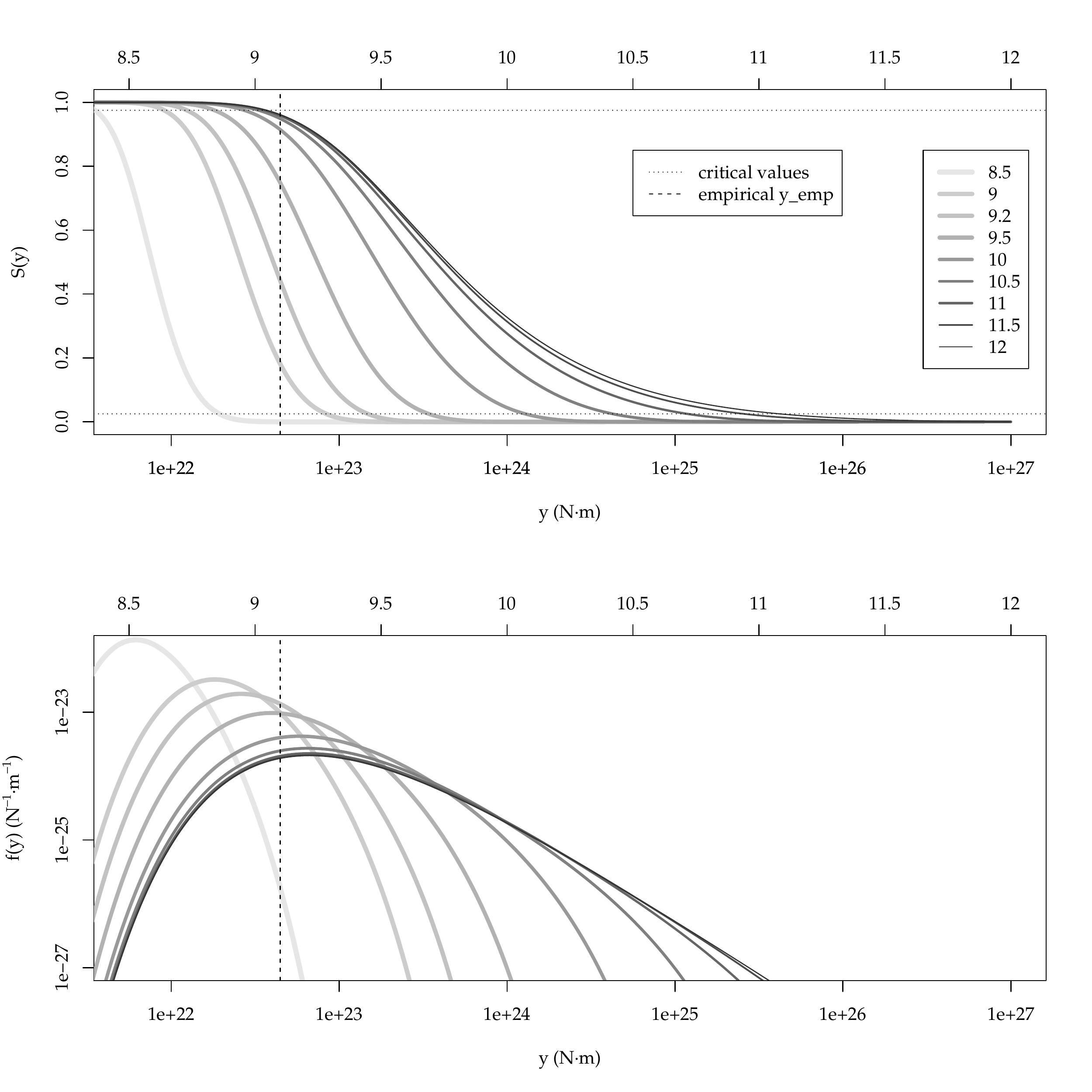}%

\caption{{\bf (Fig. S6)} ccdf $S_{max}(y)$ and pdf $f_{max}(y)$ of the maximum of 
7,585 TrG observations with
$\beta=0.67$, $a$ corresponing to moment magnitude $5.75$, and $M_c$ corresponding to the values of $m_c$ shown in the legend.
Critical values at the 95\% confidence level are shown as horizontal lines.
Empirical value of maximum seismic moment observed is shown as 
a vertical line.
}
\end{figure}


\end{document}